\pdfoutput=1

\documentclass{PoS}

\let\subsection\undefined

\newcommand{\subsection}[1]{%
  \pagebreak[2]
  \refstepcounter{subsection}
  \addcontentsline{toc}{subsection}{
    {\protect\makebox[0.3in][r]{\thesubsection} \hspace*{3pt}#1}}
  \noindent
  \textbf{{#1}\hspace{3pt}  --}
}

\usepackage{dirtree}

\DTsetlength{0.0em}{0.1em}{0.3em}{0.5pt}{2pt}
\usepackage{graphicx}
\usepackage[font=footnotesize,labelfont=bf]{caption}
\usepackage[font=footnotesize,labelfont=bf]{subcaption}

\usepackage{wrapfig}

\usepackage{units}

\def\cyp{a}
\def\ors{b}
\def\val{c}
\def\ber{d}
\def\rmii{e}
\def\fer{f}
\def\mnz{g}
\def\nic{h}
\def\hum{i}
\def\gre{j}
\def\des{k}
\def\fra{l}
\def\tre{m}
\def\jsc{n}
\def\bon{o}

\graphicspath{{graphics/}}

\title{A first look at maximally twisted mass lattice QCD calculations at the physical point}

\ShortTitle{A first look at mtmLQCD calculations at the physical point}

\def\cyp{a}
\def\ors{b}
\def\val{c}
\def\ber{d}
\def\rmii{e}
\def\fer{f}
\def\mnz{g}
\def\nic{h}
\def\hum{i}
\def\gre{j}
\def\des{k}
\def\fra{l}
\def\tre{m}
\def\jsc{n}
\def\bon{o}

\author{ A.~Abdel-Rehim$^{(\cyp)}$, 
	  Ph.~Boucaud$^{(\ors)}$,
	  N.~Carrasco$^{(\val)}$,
	  A.~Deuzeman$^{(\ber)}$,
      P.~Dimopoulos$^{(\rmii,\fer)}$,
      R.~Frezzotti$^{(\rmii)}$,
      G.~Herdoiza$^{(\mnz)}$,
      K.~Jansen$^{(\nic)}$,
      \speaker{B.~Kostrzewa}$^{(\hum,\nic)}$,
      M.~Mangin-Brinet$^{(\gre)}$,
      I.~Montvay$^{(\des)}$,
      D.~Palao$^{(\fra)}$,
      G.C.~Rossi$^{(\rmii)}$,
      F.~Sanfilippo$^{(\ors)}$,
      L.~Scorzato$^{(\tre)}$,
      A.~Shindler$^{(\jsc)}$,
      C.~Urbach$^{(\bon)}$,
      U.~Wenger$^{(\ber)}$ \\ 

      $^{(\cyp)}$ CaSToRC, The Cyprus Institute, 2121 Aglantzia, Nicosia, Cyprus

      $^{(\ors)}$ Laboratoire de Physique Th\'eorique, Universit\'e de Paris XI, 91405 Orsay-Cedex, France

      $^{(\val)}$ Departamento de F\'{\i}sica Te\`orica and IFIC, Univ. de Val\`encia-CSIC, E-46100 Val\`encia, Spain       

      $^{(\ber)}$ Albert Einstein Center for Fund. Physics, University of Bern, CH-3012 Bern, Switzerland 

      $^{(\rmii)}$ Dip. di Fisica, Universit{\`a} di Roma Tor Vergata and INFN, I-00133 Roma, Italy

      $^{(\fer)}$ Centro Fermi, Piazza del Viminale 1 I-00184 Rome, Italy            

      $^{(\mnz)}$ Institut f\"ur Kernphysik, Johannes Gutenberg-Universit\"at, D-55099 Mainz, Germany

      $^{(\nic)}$ NIC, DESY, Zeuthen, Platanenallee 6, D-15738 Zeuthen, Germany

      $^{(\hum)}$ Humboldt Universit\"at zu Berlin, Institut f\"ur Physik, Newstonstr. 15, 12489 Berlin, Germany
      
      $^{(\gre)}$ Theory Group, Lab. de Physique Subatomique et de Cosmologie, 38026 Grenoble, France

      $^{(\des)}$ Deutsches Elektronen-Synchrotron DESY, Notkestr.\,85, D-22607 Hamburg, Germany

      $^{(\fra)}$ Goethe-Universit\"at, Institut f\"ur Theoretische Physik, D-60438 Frankfurt am Main, Germany

      $^{(\tre)}$ INFN -TIFPA,  via Sommarive 14 - 38123 Trento, Italy

      $^{(\jsc)}$ IAS, IKP and JCHP, Forschungszentrum J\"ulich, 52428 J\"ulich, Germany 

      $^{(\bon)}$ HISKP (Theory), Rheinische Friedrich-Wilhelms Universit\"at Bonn, Germany
      
} % author

\abstract{In this contribution, a first look at simulations using maximally twisted mass Wilson fermions at the physical point is presented.
A lattice action including clover and twisted mass terms is presented and the Monte Carlo histories of one run with two mass-degenerate flavours at a single lattice spacing are shown.
Measurements from the light and heavy-light pseudoscalar sectors are compared to previous $N_f=2$ results and their phenomenological values.
Finally, the strategy for extending simulations to $N_f=2+1+1$ is outlined.

          \footnotesize{HU-EP-13/61, DESY 13-218, SFB/CPP-13-92} }

\FullConference{31st International Symposium on Lattice Field Theory - LATTICE 2013\\
		July 29 - August 3, 2013\\
		Mainz, Germany}

\begin{document}

\section{The Setup for Physical Point Calculations}

Performing lattice QCD calculations at the physical value of the pion mass offers the exciting possibility of avoiding the rather severe systematic uncertainties related to chiral extrapolations.
In fact, in particular in the nucleon sector, extrapolations to the physical pion mass belong to the dominant sources of systematic error. 
Simulations directly at the physical point will allow to address present discrepancies between experimental/phenomenologically extracted results and lattice QCD data in a novel way and allow to shift the focus to other systematic uncertainties as sources of these discrepancies. 
 
\begin{wraptable}{r}{0.3\textwidth}
	\centering
	\footnotesize
\begin{tabular}{|rl|}
	\hline
	L/a & 48 \\
	T/a & 96 \\
	$\beta$ & 2.10 \\ 
	$\kappa$ & 0.13729 \\
	$a\mu_l$ & 0.0009 \\
	$a\mu^\textrm{(val)}_s$ & 0.0245, 0.0252 \\
	$a\mu^\textrm{(val)}_c$ & 0.2940, 0.3058 \\
	$c_\textrm{sw}$ & 1.57551 \\
	$N_\textrm{traj}$ & $> 2000$ \\
	\hline
	$P(acc)$ & $\sim 0.75$\\
	$\langle P\rangle$ & 0.603531(6) \\
	$\tau_{\textrm{int}}(\langle P\rangle)$ & 10.0(3.5) \\ 
	$am_\textrm{PCAC}$ & 0.00004(2) \\
	$m_\pi L$ & 3.00(2) \\
	$a $ & 0.091(5) fm \\
	$r_0/a$ & $\sim 5.3$ \\
	\hline
\end{tabular}
	\caption{Run parameters and the values of the valence strange and charm quark masses. In addition, preliminary measurements of the auto-correlation time of the plaquette, the PCAC quark mass, the pion mass (in lattice units), the lattice spacing and the Sommer scale.}
	\label{tab:run_params}
\end{wraptable}

For this reason, the European Twisted Mass Collaboration (ETMC) has been exploring different gauge and fermion actions which would allow simulations directly at the physical point.
For Wilson twisted mass fermions \cite{Frezzotti:2000nk,Frezzotti:2003ni}, there exists a peculiar $\mathcal{O}(a^2)$ lattice artefact which influences the phase structure of the lattice theory \cite{Aoki:1984qi,Sharpe:1998xm,Farchioni:2004us} and leads to instabilities in the numerical simulations.
The size and nature of this lattice artefact can be parametrized by the value (and sign) of the low energy constant (LEC) $c_2$ \cite{Sharpe:1998xm}, which is related to the splitting between neutral and charged pion mass.  

One consequence of a large absolute value of $c_2$ is that simulations at the physical point can only be performed at fine lattice spacings at an unacceptable computational cost.
This turned out to be the case for the current $N_f=2+1+1$  action used by the ETMC~\cite{Baron:2010bv}.

A suitable lattice action should therefore allow for simulations at the physical point at reasonably coarse lattice spacings of around 0.1 fm.
It should further maintain all the nice properties of twisted mass lattice QCD (tmLQCD) and demonstrate control of the aforementioned $\mathcal{O}(a^2)$ discretization artefacts.

As a candidate which seems to fulfil all of these requirements, the ETMC decided to test the following action for a mass-degenerate doublet of quarks,
\begin{equation}
		\begin{array}{lll} 
			S & = \beta \sum\limits_{x;P} \left[ b_0 \lbrace 1 - \frac{1}{3} \textrm{ReTr} P^{1\times1}(x) \rbrace
					 + b_1 \lbrace 1 - \frac{1}{3} \textrm{ReTr} P^{1\times2}(x) \rbrace \right] \\
			 & + \; \sum\limits_{x} \bar{\chi}(x) \left[ D_W(U) + m_0 + i \mu \gamma^5 \tau^3 + 
			 		\frac{i}{4} c_{\textrm{sw}} \sigma^{\mu\nu} \mathcal{F}^{\mu\nu}(U)  \right] \chi(x) 
		\end{array}
		\label{eqn:action}
\end{equation}
where $b_0=1-8b_0$ and $b_1=-0.331$ are chosen to produce the "Iwasaki" gauge action \cite{Iwasaki:1983ck}.
The new ingredient compared to previous ETMC simulations is the clover-term $\frac{i}{4} c_{\textrm{sw}} \sigma^{\mu\nu} \mathcal{F}^{\mu\nu}(U)$, where $c_\textrm{sw}$ is the so-called Sheikoleslami-Wohlert improvement coefficient \cite{Sheikholeslami:1985ij}. 
The results given in this contribution were obtained using $N_f=2$ mass-degenerate twisted mass fermions tuned to maximal twist and the improvement coefficient was set to $c_\textrm{sw}=1.57551$ from Pad\'e fits to data produced by the CP-PACS/JLQCD collaboration \cite{Aoki:2005et}.
The gauge coupling parameter $\beta$ was also chosen from this data to produce a lattice spacing of roughly 0.1 fm.
The run parameters and a very preliminary determination of the lattice spacing from the pion decay constant are given in table \ref{tab:run_params}.
For computational efficiency reasons, a moderate acceptance rate of 75\% was used, with a high acceptance test run currently in progress to assess possible ill effects.
The table also lists values for the strange and charm valence quark masses that were used for measurements in the heavy-light pseudoscalar sector.
As an input, the quark mass ratios $\nicefrac{m_s}{m_l}=28$ and $\nicefrac{m_c}{m_s}=12.13$ were inferred from the corresponding values in \cite{2013:FlagReview}.
The latter differs from the FLAG value in order to approach the PDG \cite{2012:PDG} values of $\nicefrac{M_{D_s}}{f_\pi}$, $\nicefrac{M_D}{f_K}$ and $\nicefrac{M_{D_s}}{f_K}$.

Significant development effort was invested into the tmLQCD software suite \cite{Jansen:2009tmlqcd} in the form of new 'monomials' for simulations and substantial performance tuning for IBM BlueGene/Q machines.
An overview of this work was presented at this conference \cite{2013:urbach_lat13,2013:openmp_lat13}.

\section{Tuning and Stability}

\begin{figure}
	\includegraphics[width=\textwidth]{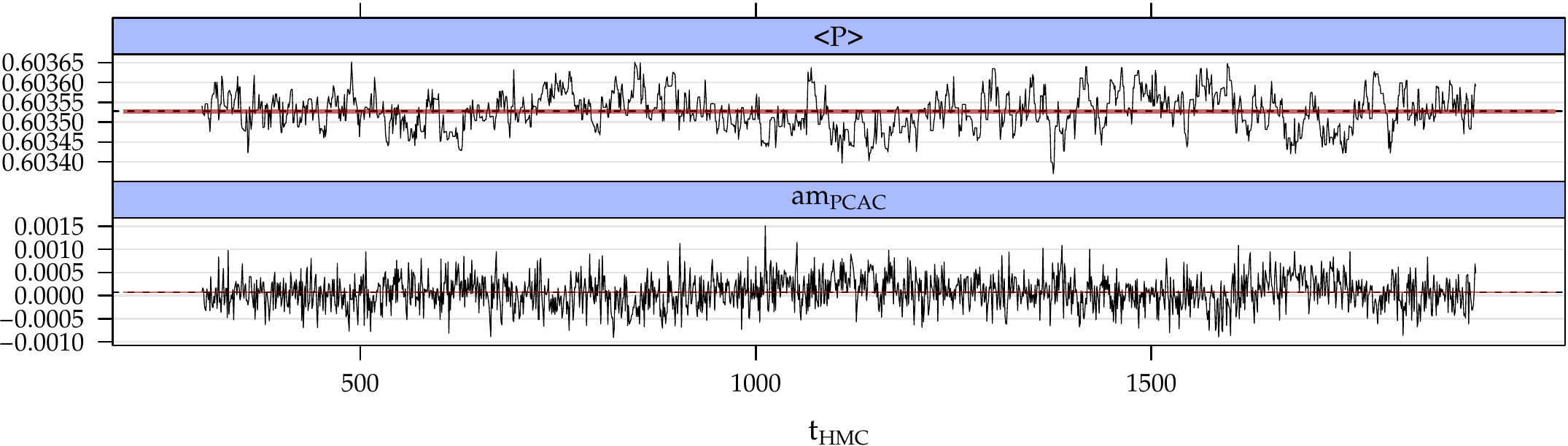}
	\caption{Monte Carlo histories of the plaquette and the PCAC quark mass.}
	\label{fig:algo}
\end{figure}

\begin{wrapfigure}{r}{0.5\textwidth}
		\centering
		\includegraphics[width=0.49\textwidth,page=7]{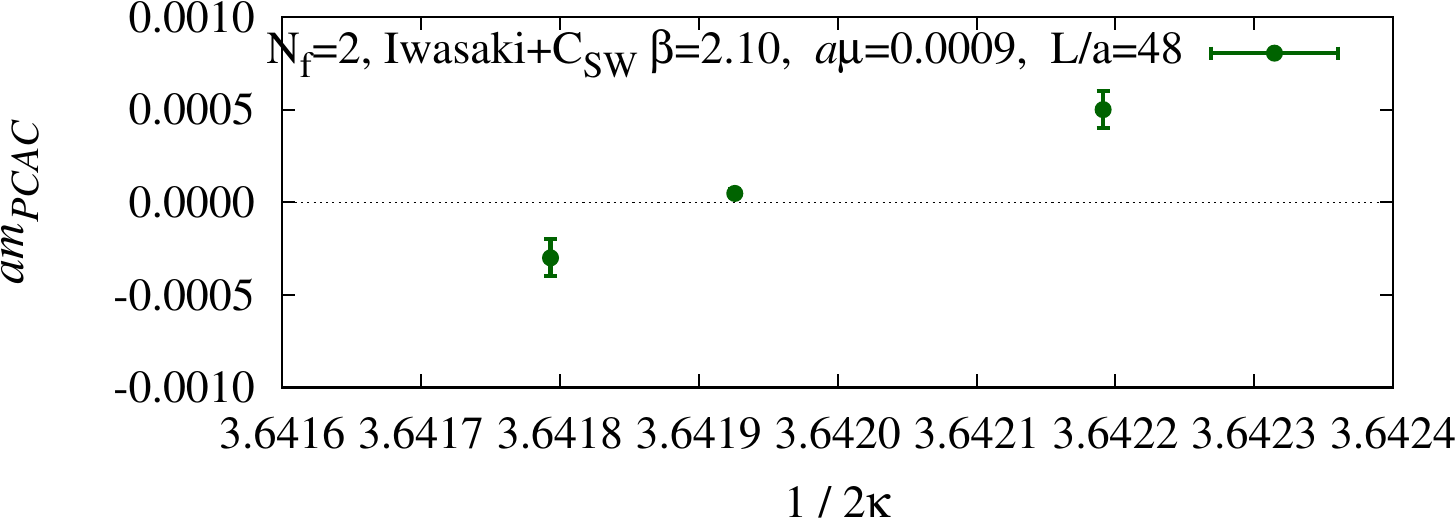}
		\caption{Behaviour of the PCAC quark mass as a function of $\nicefrac{1}{2\kappa}$ at the physical point.}
		\label{fig:pcac_vs_kappa}
\end{wrapfigure}

Tuning the hopping parameter, $\kappa$, to its critical value $\kappa_c$ and the twisted mass parameter $\mu$ to achieve the physical pion mass turned out to involve only moderate effort.
For the work in this contribution, this was achieved through about 10 runs on $24^3\times48$ lattices and quark masses $a\mu=0.006$, $a\mu=0.003$ and finally $a\mu=0.0009$.
The value of $\kappa_c$ obtained in this way had to be retuned slightly on the target volume, but the behaviour of the PCAC mass as a function of $\nicefrac{1}{2\kappa}$ has been observed to be very linear, allowing simple linear interpolations.

Monte Carlo histories of the plaquette and PCAC quark mass are shown in figure \ref{fig:algo}, while the residual tuning on the target volume is shown in figure \ref{fig:pcac_vs_kappa}.
Hot and cold starts did not point to any form of instability and topological quantities have not been computed yet.

\section{First Results}

\begin{figure}
	\begin{subfigure}{0.48\textwidth}
	\centering
			\includegraphics[height=6cm]{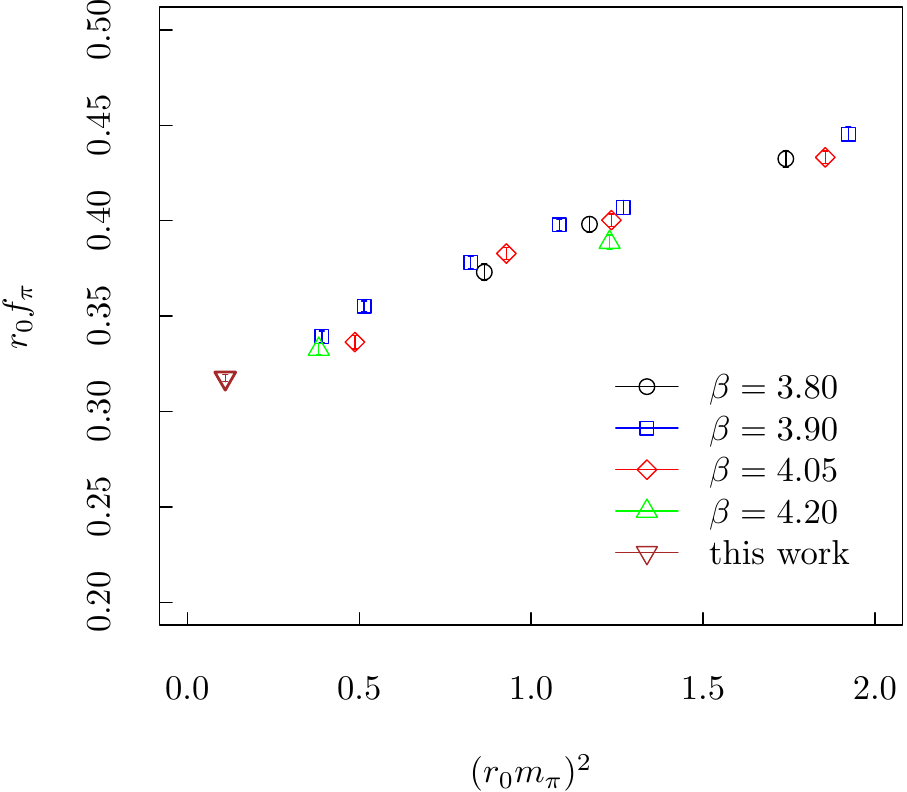}
			\caption{Pion decay constant in units of $r_0$ as a function of the squared pion mass for old ETMC $N_f=2$ data and the simulation at the physical point.}
			\label{fig:r0f_pi}
	\end{subfigure}
	\hspace{0.02\textwidth}
	\begin{subfigure}{0.48\textwidth}
		\centering
			\includegraphics[height=6cm]{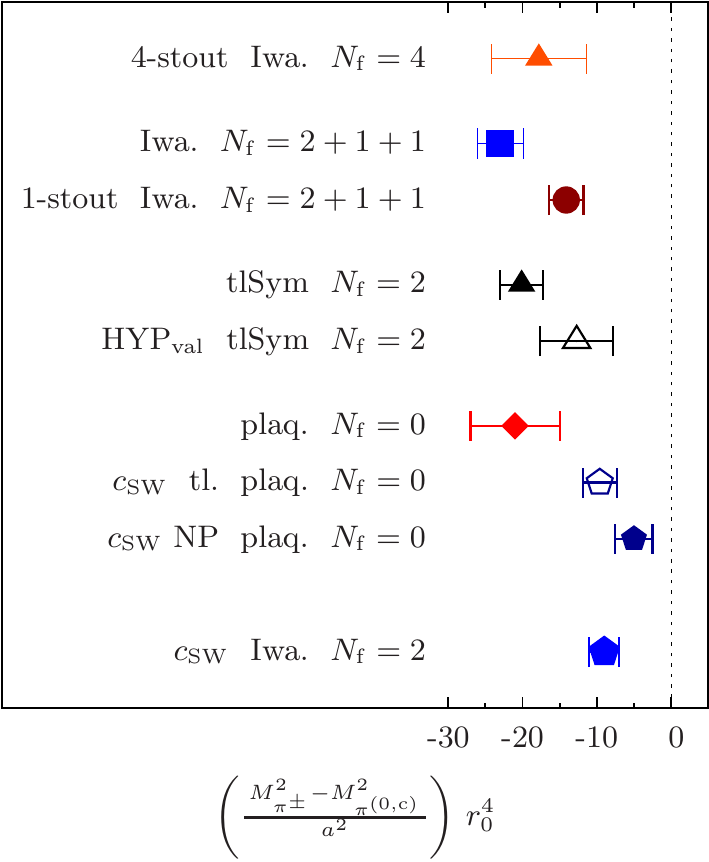}
		\caption{Mass splitting between the physical charged pion and the unphysical 'connected neutral' pion for different types of actions.
		Present work: bottom-most point}
		\label{fig:pion_split}
	\end{subfigure}
	\caption{Measurements from the pion sector compared to old results. All errors are statistical only.}
\end{figure}

Measurements of the pion decay constant in units of $r_0$ for the simulation at the physical point compared to old $N_f=2$ ETMC data are shown in figure \ref{fig:r0f_pi}.
No finite-size or lattice artefact corrections have been applied to this data and the measurement at the physical point stems from around 600 configurations.
Very preliminary determinations of renormalization constants were carried out, indicating that they are closer to their tree-level values compared to old $N_f=2$ simulations.

{\bf Isospin Breaking} -- The $\mathcal{O}(a^2)$ lattice artefacts connected to the Sharpe-Singleton scenario appear in chiral perturbation theory through the LEC $c_2$.
The calculation of the neutral pion mass involves the evaluation of disconnected diagrams needing substantial statistics which is not yet available. 
In the following, the focus will thus be on the connected part of the neutral pion mass which, however, already provides important hints for the size of $\mathcal{O}(a^2)$ lattice artefacts.

More specifically, the mass splitting in tmLQCD at non-zero lattice spacing between the charged pion and the 'neutral connected' pion can be related to the LEC $W'_8$.
In figure \ref{fig:pion_split}, a comparison between measurements of this splitting is shown for different lattice actions with different numbers of active flavours in the sea.
Clearly, the new action gives the smallest splitting amongst all actions with dynamical flavours, indicating that this lattice artefact is under much better control.
It has to be noted, however, that the splitting depends on the number of flavours and is usually worsened with increasing $N_f$.
Further, a complete understanding of the size of these $\mathcal{O}(a^2)$ artefacts is only possible through the evaluation of the full neutral pion.
A more detailed discussion can be found in \cite{2008:Shindler} and for a recent determination for previous simulations, see \cite{Herdoiza:2013sla}.

\begin{figure}
	\begin{subfigure}{0.48\textwidth}
		\centering
			\includegraphics[height=5.9cm]{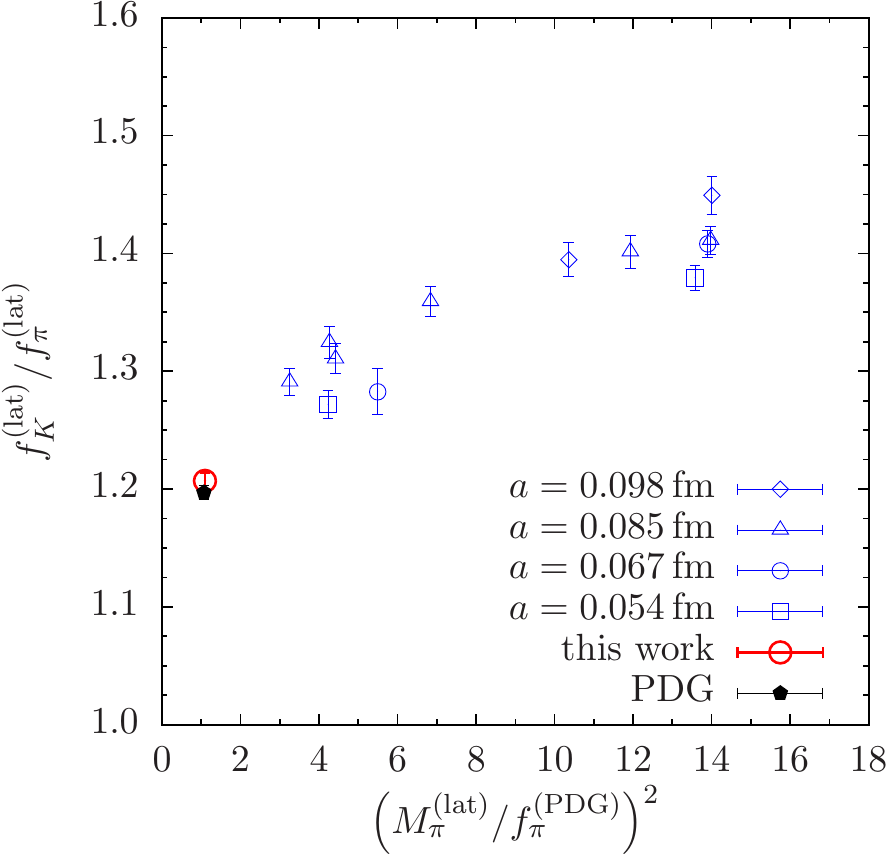}
	\end{subfigure}
	\hspace{0.02\textwidth}
	\begin{subfigure}{0.48\textwidth}
	\centering
			\includegraphics[height=5.9cm]{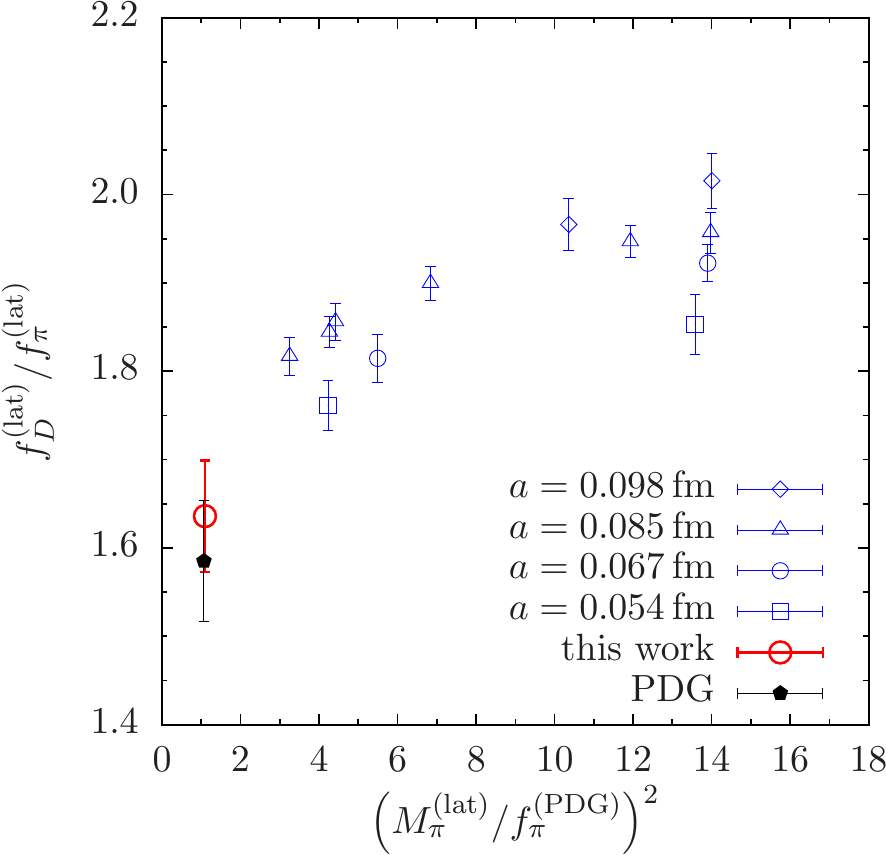}
	\end{subfigure}
	
	\begin{subfigure}{0.48\textwidth}
	\centering
			\includegraphics[height=5.9cm]{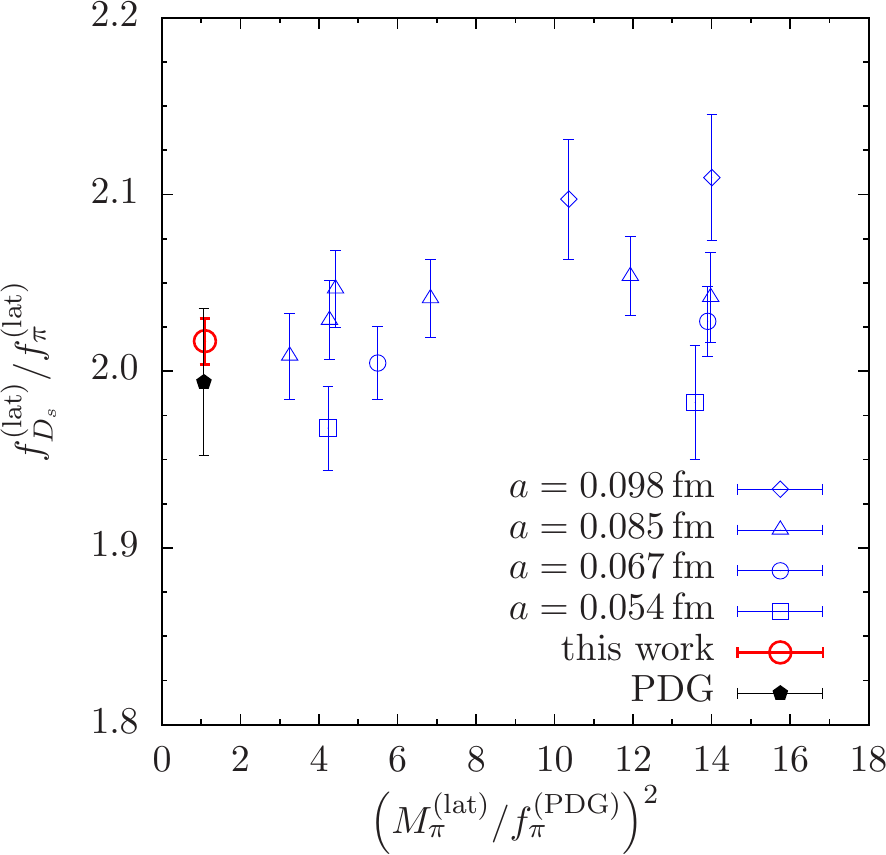}
	\end{subfigure}
	\hspace{0.02\textwidth}
	\begin{subfigure}{0.48\textwidth}
	\centering
			\includegraphics[height=5.9cm]{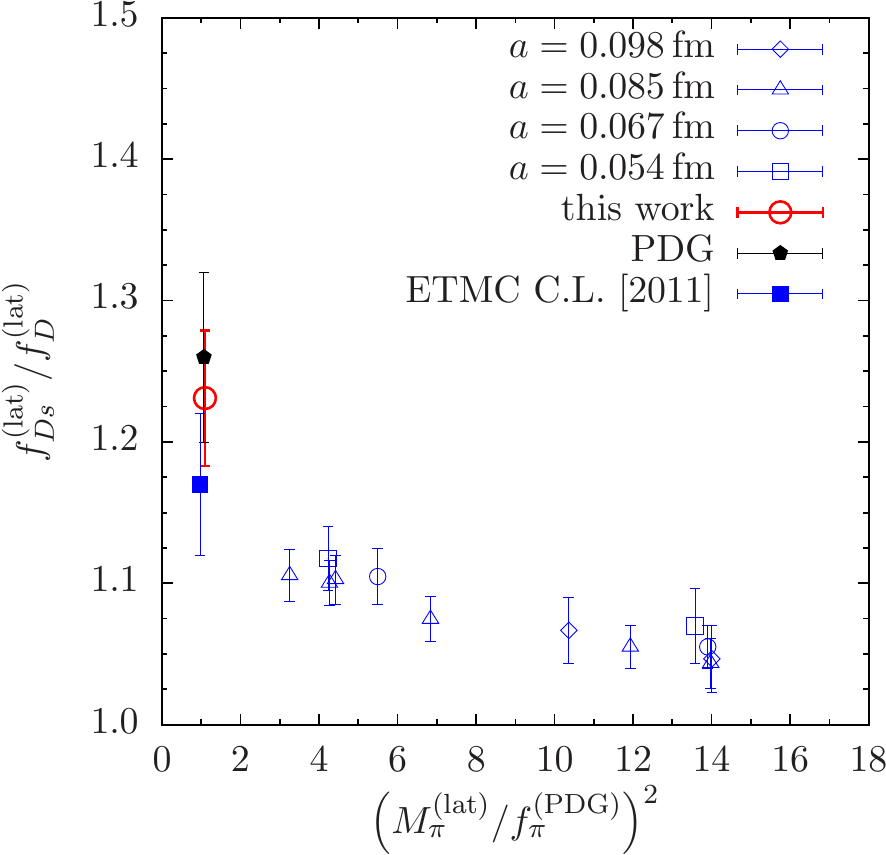}
	\end{subfigure}
	\caption{Measurements from the heavy-light pseudoscalar sector as indicated.
	The blue points correspond to old ETMC $N_f=2$ simulation results while the values from the physical point simulation are shown in red. 
	The experimental value is indicated by the black filled circle.
	For $\nicefrac{f_{D_s}}{f_D}$, the value of the chiral extrapolation of the old results is given by the filled square.
	The horizontal axis is given by the pion mass squared on the lattice in physical units, normalized by the physical value of the pion decay constant.
	All errors are statistical only with auto-correlations taken into account.}
	\label{fig:heavy-light}
\end{figure}

{\bf Heavy-light pseudoscalars} -- Determinations of the K, D and $\mathrm{D_s}$ meson masses and decay constants were carried out on 194 configurations with two values of the strange and charm quark mass as detailed in table \ref{tab:run_params}.
Decay constant ratios for the heavier masses are shown in figure \ref{fig:heavy-light}, compared to old ETMC $N_f=2$ data.
All of the measurements at the physical point coincide with their experimental counterparts and consistency is seen with the old data (keeping in mind finite-size and lattice artefact corrections, neither of which have been applied).
For $\nicefrac{f_\mathrm{D_s}}{f_\mathrm{D}}$, the chiral limit for the old data is indicated by the filled square.

Further preliminary results were presented at this conference for the nucleon sector \cite{2013:alexandrou_nucleon} and the anomalous magnetic moment of the muon \cite{2013:hotzel_gm2}.

\section{The Path to $N_f=2+1+1$}

Simulations with $N_f=2+1+1$ flavours of dynamical quarks at the physical point are clearly the eventual goal, for which the present contribution is an important proof of principle.
There are two complications which need to be addressed for this goal to be achieved.

Firstly, the stability of simulations in tmLQCD is known to be dependent on the number of dynamical flavours and the action could still turn out to be unstable at the coarse lattice spacings envisioned.
To ensure stability, runs using $N_f=2+2$, e.g. two mass-degenerate light quarks and two mass-degenerate 'strange' quarks, will be carried out.
If these simulations are stable, so will the $N_f=2+1+1$ ones for physical measurements and the $N_f=4$ ones for the determination of renormalization constants.
The $N_f=2+2$ situation has the benefit of being easier to tune than proceeding with $N_f=2+1+1$ directly and is more closely related to the $N_f=4$ situation, which is usually the least stable.

The second issue concerns the fact that there are currently no non-perturbative evaluations of the clover improvement coefficient with four dynamical flavours and Iwasaki gauge action.
Since automatic $\mathcal{O}(a)$ improvement is provided by tmLQCD at maximal twist, a fully non-perturbative value for $c_\textrm{sw}$ is not required.
Further, even non-perturbative determinations suffer from an intrinsic $\mathcal{O}(a\Lambda_\textrm{QCD})$ uncertainty, hinting at the possibility that a value within 10\% of the true non-perturbative result is sufficient for the purpose of stabilizing simulations.

\begin{wrapfigure}{r}{0.5\textwidth}
	\centering
	\includegraphics[width=0.45\textwidth]{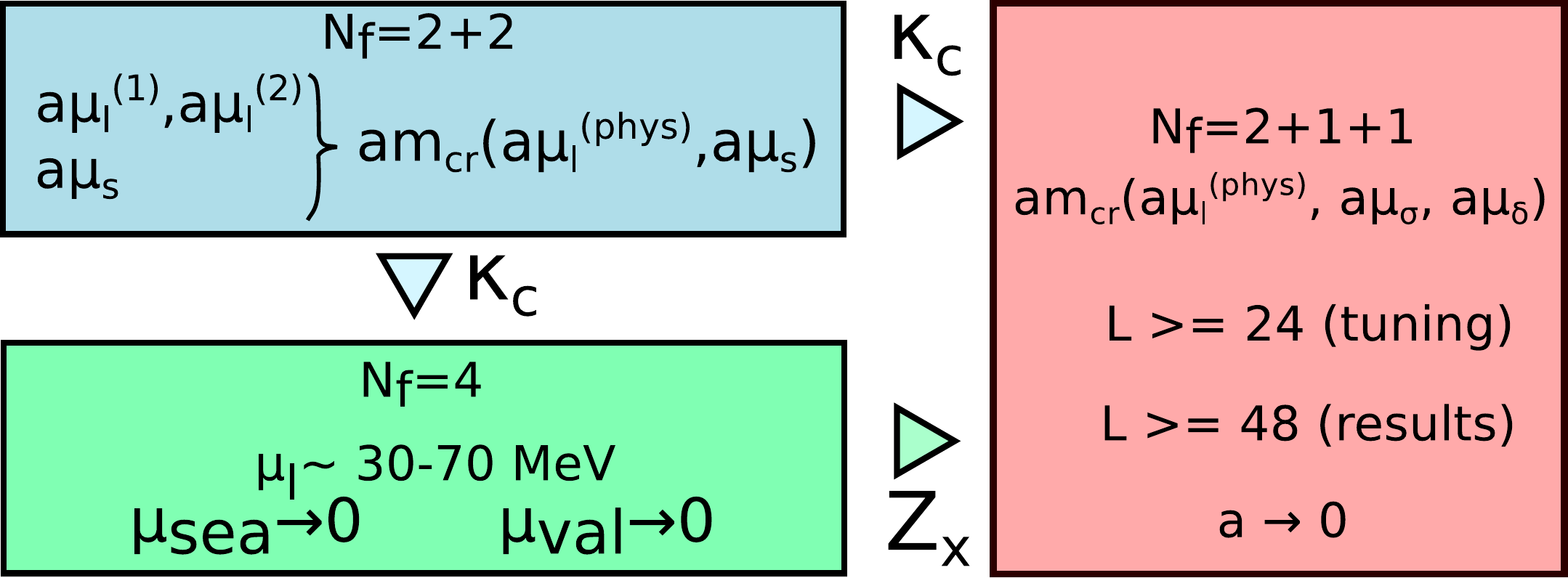}
	\caption{Pictorial representation of the information flow in the $N_f=2+2$, $N_f=4$ and $N_f=2+1+1$ simulation effort.
	$N_f=2+2$ runs demonstrate stability and extrapolations in a small number of pion masses give $\kappa_c$ at the physical point.
	The $N_f=4$ simulations at a number of quark mass values provide renormalization constants which are necessary for tuning the heavy sector in the $N_f=2+1+1$ simulations. }
	\label{fig:Nf2_Nf4_Nf211}
\end{wrapfigure}
It is planned, therefore, to use a simple tadpole-improved value \cite{Parisi:1980tadpole,1999:csw_frezzotti} of the clover coefficient
\begin{equation}
  c_\textrm{sw} \sim 1 + 0.113(3) \frac{g_0^2}{\langle P\rangle}\,,
  \label{eqn:csw_tadpole}
\end{equation}
where $g_0$ is the bare coupling at the given lattice spacing and $\langle P\rangle$ is the plaquette expectation value.

\begin{wrapfigure}{r}{0.5\textwidth}
	\centering
	\includegraphics[width=0.5\textwidth,page=8]{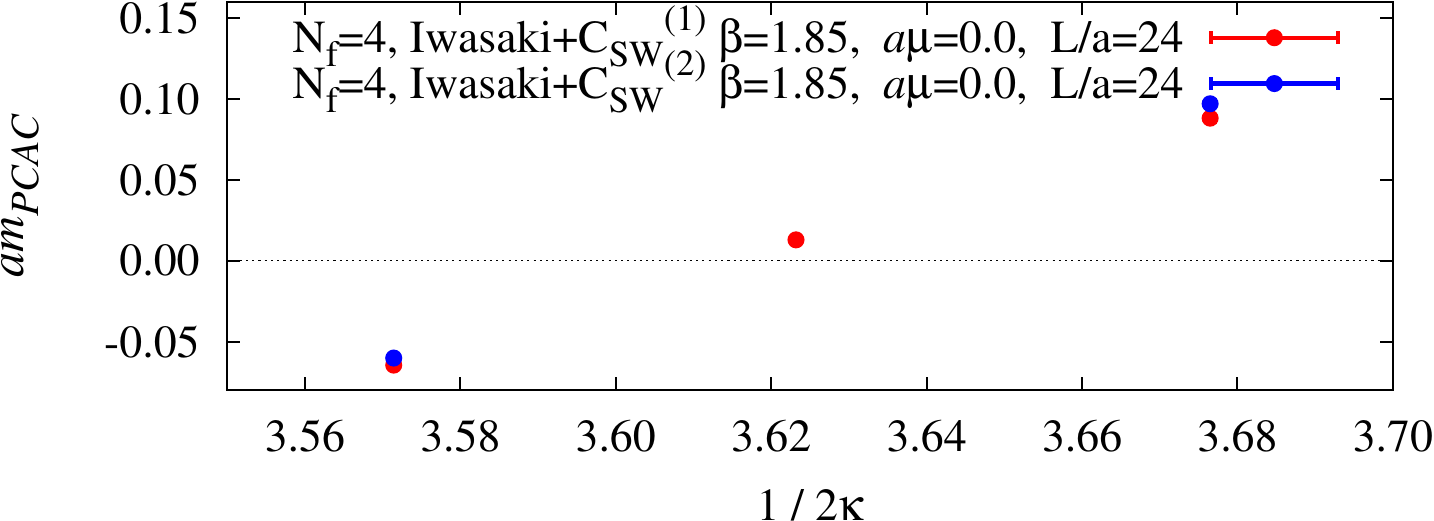}
	\caption{Behaviour of $am_\textrm{PCAC}$ as a function of $\nicefrac{1}{2\kappa}$ for $N_f=4$ runs with two different estimates of $c_\textrm{sw}$. In these simulations the twisted quark mass $a\mu$ has been set to 0.}
	\label{fig:tuning_csw}
\end{wrapfigure}

Interestingly, the current value of 1.57551 lies within 3\% of the value obtained from equation \ref{eqn:csw_tadpole}.
For new $N_f=2+1+1$ simulations, $c_\textrm{sw}$ can be obtained by a simple algorithm, starting at some first guess $c^0_\textrm{sw}$. 
The $\kappa$ tuning runs on small volumes can be exploited by interpolating the plaquette expectation value for negative and positive PCAC quark masses to the critical point and plugging it into the above equation.
This value of $c_\textrm{sw}$ can then be used for further runs, repeating the process of measuring $\langle P \rangle$ and updating $c_\textrm{sw}$, until it remains stable within, say, 3\% of the previous value.
Tuning to $\kappa_c$ in all $N_f=2+2$, $N_f=4$ and $N_f=2+1+1$ runs can then be continued with this final estimate.

Renormalization constants can be computed in the massless limit from $N_f=4$ simulations on smaller lattices with a range of unphysically heavy quark masses.
These can then inform the tuning effort in the heavy sector of the $N_f=2+1+1$ simulations on small volumes.
With all parameters tuned appropriately, large volume simulations can finally be performed with minimal retuning.
A pictorial representation of the interplay of the different runs is shown in figure \ref{fig:Nf2_Nf4_Nf211}.

First runs with four mass-degenerate flavours have already been attempted at $\beta=1.85$.
The algorithm for tuning $c_\textrm{sw}$ has been tested and shown to work well even with vanishing twisted quark mass $a\mu$ as shown in figure \ref{fig:tuning_csw}.
It appears, however, that the lattice spacing is quite fine at this value of the gauge coupling parameter and tuning is currently on-going for coarser lattice spacings.

\section{Conclusion and Outlook}

In this contribution, preliminary measurements in the light and heavy-light pseudoscalar sectors from a simulation directly at the physical point using two flavours of mass-degenerate twisted mass quarks at maximal twist have been presented.
It has been shown that for this situation, the new lattice action using an "Iwasaki" gauge and a clover term seems to yield stable simulations.
Measurements of decay constants in the heavy-light pseudoscalar meson sector agree very well with their experimental values with minimal tuning effort of the valence strange and charm quark masses.
In the near future, simulations will be extended to $N_f=2+1+1$ flavours with multiple lattice spacings and increasing volumes.

\section*{Acknowledgements}

B.K. acknowledges full support by the National Research Fund, Luxembourg under AFR Ph.D. grant 27773315. 
L.S. acknowledges partial support from the INFN - SUMA project and the EU STRONGnet project.
A.A.-R. acknowledges support from the PRACE 2IP project under grant number EC-RI-283493.
D.P. acknowledges the HIC for FAIR within the framework of the LOEWE program launched by the State of Hesse for partial support.
This work is supported in part by DFG and NSFC (CRC 110).
The computations for this work were carried out on JuQueen, Judge and Juropa at JSC Juelich, SuperMUC at LRZ Munich and Fermi at Cineca Bologna under grants provided through PRACE, the agreement between INFN and CINECA under initiative INFN-RM123 and the Gauss center for supercomputing.

\bibliographystyle{3authors_notitle}
\bibliography{bibliography}{}

\begin{thebibliography}{10}
\providecommand{\url}[1]{\texttt{#1}}
\providecommand{\urlprefix}{URL }
\providecommand{\eprint}[2][]{\url{#2}}

\bibitem{Frezzotti:2000nk}
R.~Frezzotti \emph{et~al.} (ALPHA).
\newblock JHEP (2001).
\newblock 08:058.
\newblock \eprint{hep-lat/0101001}

\bibitem{Frezzotti:2003ni}
R.~Frezzotti and G.~C. Rossi.
\newblock JHEP (2004).
\newblock 08:007.
\newblock \eprint{hep-lat/0306014}

\bibitem{Aoki:1984qi}
S.~Aoki.
\newblock Phys.Rev. (1984).
\newblock D30:2653

\bibitem{Sharpe:1998xm}
S.~R. Sharpe and J.~Singleton, R.
\newblock Phys. Rev. (1998).
\newblock D58:074501.
\newblock \eprint{hep-lat/9804028}

\bibitem{Farchioni:2004us}
F.~Farchioni \emph{et~al.}
\newblock Eur. Phys. J. (2005).
\newblock C39:421--433.
\newblock \eprint{hep-lat/0406039}

\bibitem{Baron:2010bv}
R.~Baron \emph{et~al.} (ETM).
\newblock JHEP (2010).
\newblock 1006:111.
\newblock \eprint{1004.5284}

\bibitem{Iwasaki:1983ck}
Y.~Iwasaki.
\newblock UTHEP-118

\bibitem{Sheikholeslami:1985ij}
B.~Sheikholeslami and R.~Wohlert.
\newblock Nucl. Phys. (1985).
\newblock B259:572

\bibitem{Aoki:2005et}
S.~Aoki \emph{et~al.} (CP-PACS).
\newblock Phys. Rev. (2006).
\newblock D73:034501.
\newblock \eprint{hep-lat/0508031}

\bibitem{2013:FlagReview}
S.~Aoki \emph{et~al.} (FLAG) (2013).
\newblock \eprint{1310.8555}

\bibitem{2012:PDG}
J.~Beringer \emph{et~al.} (Particle Data Group).
\newblock Phys. Rev. D (2012).
\newblock 86:010001

\bibitem{Jansen:2009tmlqcd}
K.~Jansen and C.~Urbach.
\newblock Comput.Phys.Commun. (2009).
\newblock 180:2717--2738.
\newblock \eprint{0905.3331}

\bibitem{2013:urbach_lat13}
A.~Abdel-Rehim \emph{et~al.} (ETM).
\newblock In \emph{31st International Symposium on Lattice Field Theory}, no.
  414 in PoS(LATTICE 2013)

\bibitem{2013:openmp_lat13}
A.~Deuzeman \emph{et~al.} (ETM).
\newblock In \emph{31st International Symposium on Lattice Field Theory}, no.
  416 in PoS(LATTICE 2013)

\bibitem{2008:Shindler}
A.~Shindler.
\newblock Physics Reports (2008).
\newblock 461(2-3):37--110.
\newblock \eprint{0707.4093}

\bibitem{Herdoiza:2013sla}
G.~Herdoiza \emph{et~al.} (ETM).
\newblock JHEP (2013).
\newblock 1305:038.
\newblock \eprint{1303.3516}

\bibitem{2013:alexandrou_nucleon}
C.~Alexandrou \emph{et~al.} (ETM).
\newblock In \emph{31st International Symposium on Lattice Field Theory}, no.
  292 in PoS(LATTICE 2013)

\bibitem{2013:hotzel_gm2}
F.~Burger \emph{et~al.} (ETM).
\newblock In \emph{31st International Symposium on Lattice Field Theory}, no.
  301 in PoS(LATTICE 2013)

\bibitem{Parisi:1980tadpole}
G.~Parisi.
\newblock AIP Conference Proceedings (1980).
\newblock 68(1):1531--1568.
\newblock ICHEP

\bibitem{1999:csw_frezzotti}
S.~Aoki, R.~Frezzotti and P.~Weisz.
\newblock Nuclear Physics B (1999).
\newblock 540(1-2):501.
\newblock \eprint{hep-lat/9808007}

\end{thebibliography}

\end{document}